\documentclass[prd,twocolumn,showpacs,letterpaper,nofootinbib]{revtex4}
\usepackage{amsmath}
\usepackage{graphicx}
\usepackage{graphics} 
\usepackage{color} 

\usepackage{amssymb}

\newcommand{\be}{\begin{equation}}
\newcommand{\ee}{\end{equation}}
\newcommand{\bea}{\begin{eqnarray}}
\newcommand{\eea}{\end{eqnarray}}

\begin{document}

\title{Generalized Israel Junction Conditions for a Fourth-Order Brane World}
\author{Adam Balcerzak}
\email{abalcerz@wmf.univ.szczecin.pl}
\author{Mariusz P. D\c{a}browski}
\email{mpdabfz@sus.univ.szczecin.pl}
\affiliation{Institute of Physics, University of Szczecin,
Wielkopolska 15, 70-451 Szczecin, Poland.}

\date{\today}

\begin{abstract}
We discuss a general fourth-order theory of gravity on the brane. In general, the formulation of the junction conditions (except
for Euler characteristics such as Gauss-Bonnet term) leads to the higher powers of the delta
function and requires regularization. We suggest the way to avoid
such a problem by imposing the metric and its first derivative to
be regular at the brane, while the second derivative to have a kink, the third
derivative of the metric to have a step function discontinuity, and no sooner as the
fourth derivative of the metric to give the delta function
contribution to the field equations. Alternatively, we discuss the reduction of the fourth-order gravity
to the second-order theory by introducing an extra tensor field. We formulate the
appropriate junction conditions on the brane. We prove the equivalence of both theories.
In particular, we prove the equivalence of the junction conditions with
different assumptions related to the continuity of the metric along the brane.
\end{abstract}


\maketitle

\section{Introduction}
\label{sect1}

\setcounter{equation}{0}

Brane universes have made great popularity during the last years \cite{RS,brane}.
However, it is remarkable that so far only the standard Einstein
gravity, Gauss-Bonnet gravity \cite{deruelle00,charmousis,davis,jim,lidsey,maeda,apostopoulos} and, in general, Euler density
gravity \cite{lovelock} on the brane have been considered in the literature \cite{meissner01}. These can be
expressed by the general action \footnote{We use convention (-++...+) for the metric following Ref. \cite{he}.}
\bea
\label{euler}
S = \int_M d^D x \sqrt{-g} \sum_n \kappa_n I^{(n)} + S_{brane} + S_{m}~,
\eea
where $I^{(n)}$ is the Euler density of the n-th order, $\kappa_n$
is an appropriate constant of the n-th order, $M$ is a $D$-dimensional
manifold, $S_{brane}$ is the brane action and $S_m$ is the matter action. The lowest order
Euler densities are: the cosmological constant $I^{(0)} = 1$, the Ricci scalar $I^{(1)} = R$,
and the Gauss-Bonnet density $I^{(2)} = R_{GB} = R_{abcd}R^{abcd} - 4 R_{ab}R^{ab} +R^2$
with appropriate constants $\kappa_0 = -2\Lambda(2\kappa^2)^{-1} = -2\Lambda/16 \pi
G$, $\kappa_1 = (2\kappa^2)^{-1}$, $\kappa_2=\alpha(2\kappa^2)^{-1}$, $\alpha=$ const. etc.,
$a,b,c=0,1,\ldots, D-3, D-2, D$ \cite{davis}.

In fact, in a general class of brane models based on an arbitrary combination
of the higher-order curvature terms $f(R_{abcd}R^{abcd},R_{ab}R^{ab},R)$
the field equations are fourth-order. Because of that they are plagued by the higher power
terms of the second derivative of the warp factor function
$\sigma(y) = \mid y \mid$. This leads to a production of the higher powers of the delta function
$\delta(y)$ which can make the field equations ambiguous.
Among the general class, the models based on the Euler densities are unique
in the sense that the higher powers of the second derivative of the warp factor
$\partial^2 \sigma(y)/\partial y^2$ exactly cancel in the field equations \cite{meissner01}.
One then is easily able to formulate
appropriate junction conditions given first by Israel \cite{israel66,visser}.

The main objective of our paper will be the study of a
general {\it fourth-order theory of gravity on the brane} \cite{clifton}
\bea
\label{XYZ}
S &=& \chi^{-1} \int_{M} d^{D}x \sqrt{-g} f(X,Y,Z) + S_{brane} + S_{m}~,
\eea
where $X=R$, $Y=R_{ab}R^{ab}$, $R_{abcd}R^{abcd}$ are curvature
invariants, and $\chi$ is a constant. It includes the Euler
density theories with the first Euler density being just $f(X,Y,Z) = \chi \kappa_1 X = \chi \kappa_1 R $
and the second Euler density being the Gauss-Bonnet term given by $f(X,Y,Z) = \chi \kappa_2 (Z - 4Y +X^2)$
etc.

Up to our knowledge, the only non-Eulerian density cases were studied in Refs.
\cite{branef(R)} and \cite{braneR2}. In Ref. \cite{branef(R)} the fourth-order theory
$f(X,Y,Z) = f(X) = f(R)$ was first
reduced to the second-order theory, and then transformed into the
Einstein theory. The junction conditions were then obtained, and
they were obviously free from the problem of the powers of
$\delta-$function contribution. On the other hand, in Ref.
\cite{braneR2} the theories with the linear combination of the
form $f(X,Y,Z) = aX^2 + bY + cZ$ ($a,b,c=$ const.) were considered
and the junction conditions were obtained by the application of
the appropriate Gibbons-Hawking boundary terms, again after transforming
this theory to an equivalent second-order theory.

It is important to emphasize that the theories based on the functions of the Euler densities
such as $f(I^{(n)})$ are fourth-order. Among them the most popular are
$f(I^{(1)}) = f(R)$ - the theories of the function of the first Euler density \cite{f(R)}. In fact, the
theories which are based on the function of the second Euler density $f(I^{(2)})$ have also gained
some interest recently \cite{f(RGB)}, but they have not been studied on the brane yet.

Our paper is organized as follows.
In Section II we discuss the main obstacle to formulate junction
conditions for the fourth-order braneworld in a standard way which
has been performed in the case of Euler densities. In Section III
we make a proposal to formulate these junction conditions by
imposing more regularity onto the metric tensor. Since it does not
necessarily satisfy everybody's taste we present in Section IV an
alternative approach. In this approach we transform our general
fourth-order theory into a second-order theory by applying
generalized Lagrange-multiplier approach \cite{f(R),f(RGB),kijowski}. This method was
successful in obtaining the junction conditions in $f(R)$ theory
\cite{branef(R)} and in $f = aX^2 + bY + cZ$ theory \cite{braneR2}.
In the Section V we formulate the junction
conditions for the equivalent second-order theory. Finally, in
Section VI we give our conclusions.

\section{Problem formulating Israel junction conditions in a fourth-order brane world}
\label{sect2}

\setcounter{equation}{0}

In order to discuss the problem let us begin with
the standard D-dimensional brane theory \cite{RS} whose action is just
(\ref{euler}) with only $\kappa_1 \neq 0$, i.e.,
\bea
S = \frac{1}{2\kappa^2}\int_{M} d^{D}x \sqrt{-g} R + S_{brane} +
S_{m}
\eea
with the field equations
\bea
G_a^{~b}=R_a^{~b}-(1/2)\delta_a^{~b}R=\kappa^2 T_a^{~b} ,
\eea
where the energy-momentum tensor is given by
\begin{eqnarray}
\label{Tab}
T_a^{~b}=T_{a}^{~b~-}\Theta(-w) + T_{a}^{~b~+}\Theta(w) +
\delta(w)S_a^{~b},
\end{eqnarray}
with $S_a^{~b}$ being the energy-momentum tensor on the brane, and
$T_{a}^{~b~\pm}$ are the energy-momentum tensors on the both sides of the brane
(i.e., in the bulk).

For simplicity, let us assume that we work in Gaussian normal coordinates
- i.e., that the D-dimensional metric is of the form $(\mu,\nu = 0, 1, 2,\ldots,D-2;w=D)$
\begin{eqnarray}
\label{bm}
ds^2=g_{ab} dx^a dx^b = \epsilon dw^2+h_{\mu\nu}dx^{\mu}dx^{\nu}~,
\end{eqnarray}
where $\epsilon = \vec{n} \cdot \vec{n} = +1$ for a spacelike hypersurface,
$\epsilon= -1$ for a timelike hypersurface, and $h_{ab} = g_{ab} - \epsilon n_a n_b$
is a projection tensor onto a $(D-1)$-dimensional hypersurface, $\vec{n}$ is the normal vector to
the hypersurface. The extrinsic
curvature in these coordinates is defined as
\begin{eqnarray}
K_{\mu\nu}=-{1\over 2}{\partial h_{\mu\nu}\over\partial w}~.
\end{eqnarray}

By the application of the Gauss-Codazzi equations \cite{mannheim}
\begin{eqnarray}
\label{GC}
R_{w\mu w\nu}&=& {\partial K_{\mu\nu}\over \partial w}+K_{\rho\nu}K^{\rho}_{\,\,\mu},  \\
R_{w\mu\nu\rho}&=&\nabla_{\nu}K_{\mu\rho}-\nabla_{\rho}K_{\mu\nu}, \\
R_{\lambda\mu\nu\rho}&=&~^{(D-1)} R_{\lambda\mu\nu\rho}+
\epsilon\left[K_{\mu\nu}K_{\lambda\rho}
-K_{\mu\rho}K_{\lambda\nu}\right]~,
\end{eqnarray}
one has the D-dimensional field equations in the form
\begin{eqnarray}
\label{Gww}
G^w_{~w}&=&-{1\over 2}~^{(D-1)}R+{1\over 2}\epsilon\left[K^2-Tr(K^2)\right]=\kappa^2 T^w_{~w}, \\
\label{Gwm}
G^w_{~\mu}&=&\epsilon\left[\nabla_{\mu}K-\nabla_{\nu}K^{\nu}_{\,\,\mu}\right]=\kappa^2
T^w_{~\mu}, \\
\label{Gmm}
G^{\mu}_{~\nu}&=&~^{(D-1)}G^{\mu}_{~\nu}
             +\epsilon\left[{\partial K^{\mu}_{~\nu}\over\partial w}-\delta^{\mu}_{~\nu}
{\partial K\over\partial w}\right]\\
&+& \epsilon\left[-
K K^{\mu}_{~\nu}+{1\over 2}\delta^{\mu}_{~\nu}Tr(K^2)+{1\over 2}\delta^{\mu}_{~\nu}
K^2\right]=\kappa^2 T^{\mu}_{~\nu}~.\nonumber
\end{eqnarray}

The Israel junction conditions can be obtained by the
integration of the field equations (\ref{Gww})-(\ref{Gmm}) in the
limit $ \lim_{w \to 0} \int_{-w}^{w}$ \cite{deruelle00} and read as
\begin{eqnarray}
\label{S3}
\epsilon \{ [K^{\mu}_{~\nu}]-\delta^{\mu}_{~\nu}[K]\} &=& \kappa^2 {S}^{\mu}_{~\nu},\\
\label{S1}
0 &=& \kappa^2 {S}^w_{~w}, \\
\label{S2}
0 &=& \kappa^2 {S}^w_{~\mu},
\end{eqnarray}
where
$[K^{\mu}_{~\nu}] \equiv K^{\mu~+}_{~\nu}-K^{\mu~-}_{~\nu}$.
In general, for any quantity $\Omega$, one defines $[\Omega] = \Omega^{+} -
\Omega^{-}$, where $\Omega^{\pm}$ means that this quantity was calculated on
the left-hand-side and on the right-hand-side of the brane, respectively.

The most important point is that these junction conditions are
obtained provided we assume the following continuity conditions
for the metric at $w=0$ \cite{visser}:
\begin{eqnarray}
\label{cont1}
h^{-}_{\mu\nu} &=& h^{+}_{\mu\nu}~,\\
\label{cont2}
h^{-}_{\mu\nu,w} & \neq & h^{+}_{\mu\nu,w}~, \hspace{0.5cm} K^{-}_{\mu\nu}
\neq K^{+}_{\mu\nu}~,
\end{eqnarray}
which means that the metric is continuous at the brane but it has a kink, its first
derivative has a step function discontinuity, and its second derivative gives the delta
function contribution. In other words:
\begin{eqnarray}
h_\mu{_\nu}(w) &=& h^{-}_{\mu\nu}(w) \theta(-w) + h^{+}_{\mu\nu}(w)
\theta(w) ~,\\
{\partial h^{-}_{\mu\nu}} \over \partial w &=&  {\partial h^{+}_{\mu\nu} \over
\partial w} \theta(-w) +  {\partial h^{-}_{\mu\nu} \over \partial w} \theta(w)~, \\
{\partial{^2} {h_\mu{_\nu}} \over \partial w{^2}}&=&  {\partial{^2} h^{-}_{\mu\nu}
\over \partial w{^2}} \theta(-w) +  {\partial{^2} h^{+}_{\mu\nu} \over \partial w{^2}}
\theta(w) \nonumber \\ &+& \left( {\partial h^{-}_{\mu\nu} \over \partial w} -
{\partial h^{+}_{\mu\nu} \over \partial w} \right)\delta(w)~.
\end{eqnarray}

The equation (\ref{S3}) follows from (\ref{Gmm}) and (\ref{Tab})
as a consequence of the fact that the terms $\partial K_{\mu}^{\nu}/ \partial w$
and $\partial K / \partial w$ contain the delta function $\delta(w)$. However,
in the fourth-order theory given by the action (\ref{XYZ}), the
application of the continuity conditions (\ref{cont1})-(\ref{cont2}) does not work.
In order to discuss this let us first write down the field
equations for the action (\ref{XYZ}) \cite{clifton}:
\begin{eqnarray}
\label{XYZ1}
P_{a b}&=&\frac{\chi}{2} T_{a b}, \\
\label{XYZ2}
P^{a b} &=& -\frac{1}{2} f g^{a b} + f_X R^{a b}+2 f_Y R^{c (a} {R^{b)}}_{c}+2
f_Z R^{e d c (a} {R^{b)}}_{c d e} \nonumber \\ &+& f_{X; c d}(g^{a
b} g^{c d}-g^{a c} g^{b d}) + \Box(f_Y R^{a b}) + g^{a b} (f_Y
R^{c d})_{;c d} \nonumber \\ &-& 2 (f_Y R^{c (a})_{;\; \; c}^{\;
b)}-4 (f_Z R^{d (a b) c})_{;c d},
\end{eqnarray}
where $f_X = {\partial f / \partial X}$ etc. The reason for not being the same continuity
conditions (\ref{cont1})-(\ref{cont2}) valid here is that the Riemann tensor
\begin{eqnarray}
R&=&~^{(D-1)}R+\epsilon\left[2h^{\mu\nu}{\partial K_{\mu\nu}\over\partial w}
+3Tr(K^2)-K^2\right]~,\nonumber
\end{eqnarray}
where $K\equiv K^{\mu}_{\,\,\mu}$ , $Tr(K^2)\equiv K^{\mu\nu}K_{\mu\nu}$,
and appropriately, the Ricci tensor, and the Ricci scalar, as squared,
give the terms
\bea
{\partial^2 \gamma^{\mu \nu} \over \partial^2 w}{\partial K_{\mu \nu} \over \partial w},
   {\partial K_{\mu \nu} \over \partial w}{\partial K^{\mu \nu} \over \partial w},
   \left({\partial K \over \partial w}\right)^2~,
\eea
which are proportional to $\delta^2(w)$,
For example, in the simple fourth-order gravity theory $f(X,Y,Z)=f(X)=R^2$
the field equations contain the term ${f_X},_{ww}$, which is proportional
to $\delta^2(w)$.

As mentioned in the Introduction, the only ``mysterious'' cases which do not involve
singular expressions are the Euler densities. In these cases the terms proportional to
$\delta^2(w)$ exactly cancel. These theories have been studied by a number
of people (e.g. \cite{deruelle00,charmousis,davis,jim,lidsey,maeda}) and their field equations are given by
varying the action (\ref{euler}) with $\kappa_1 \neq 0$, $\kappa_1 \neq 0$ and $\kappa_2 \neq 0$
\begin{eqnarray}
\label{GB1}
   G_{ab}&+&\alpha H_{ab} + \Lambda g_{ab} = \kappa^2 T_{ab}, \\
\label{GB2}
   G_{ab}&\equiv& R_{ab}-{1\over2}g_{ab}R, \\
\label{GB3}
   H_{ab}&\equiv& 2(R_{almn}R_b^{\,\,\,lmn}-2R_{ambn}R^{mn} \\
   &-& \nonumber 2R_{am}R_b^{\,\,\,m}+RR_{ab})  -{1\over 2}
       g_{ab}I^{(2)}. \nonumber
\end{eqnarray}
The ``dangerous'' terms appear in the ``correction'' $H_{ab}$ to the
Einstein equations which is
\begin{eqnarray}
&& H^{\mu}_{~\nu}= 4{\partial\over\partial w}
\left\{KK^{\mu}_{~\alpha}K^{\alpha}_{~\nu}-K^{\mu}_{~\alpha}K^{\alpha\beta}
K_{\beta\nu} \right. \nonumber \\
&+& \left. {1\over 2}K^{\mu}_{~\nu}Tr(K^2)- {1\over 2}K^{\mu}_{~\nu}K^2\right\}\nonumber\\
                 &+&4{\partial\over\partial w}\left\{-\delta^{\mu}_{~\nu}{1\over 2}K
                 Tr(K^2)+\delta^{\mu}_{~\nu} {1\over 3} Tr(K^3)
                 +\delta^{\mu}_{~\nu}{1\over 6}K^3\right\}\nonumber\\
                 &+&4\left(-~^4R_{~\alpha\nu}^{\mu~~~\beta}
                 {\partial K^{\alpha}_{~\beta}\over\partial w}
                 -~^4R^{\alpha}_{~\nu}{\partial K^{\mu}_{~\alpha}\over\partial w}
                 -~^4R^{\alpha\mu}{\partial K_{\nu\alpha}\over\partial w}\right)\nonumber\\
                 &+&4\left(~^4R^{\mu}_{~\nu}{\partial K \over\partial w}
                   +{1\over 2}~^4R{\partial K^{\mu}_{~\nu}\over\partial w}
                   +\delta^{\mu}_{~\nu}~^4R^{\alpha\beta}
                   {\partial K_{\alpha\beta}\over\partial w}
                   \right. \nonumber \\
                  &-& \left. {1\over 2}\delta^{\mu}_{~\nu}~^4R{\partial K \over\partial w}
                  \right) + \ldots\,,
\end{eqnarray}
The junction conditions are obtained by using the
continuity conditions (\ref{cont1})-(\ref{cont2}) and read \cite{davis}
\bea
&& 2 \alpha \left( 3 [J_{\mu\nu}] - [J] h_{\mu\nu}
- 2 [P]_{\mu\rho\nu\sigma} [K]^{\rho\sigma} \right) \nonumber \\
&+& [K_{\mu\nu}] - [K] h_{\mu\nu} = - \kappa^2 S_{\mu\nu}~,
\eea
where
\bea
P_{\mu\rho\nu\sigma} &=& R_{\mu\rho\nu\sigma} + 2 h_{\mu[\sigma}R_{\nu]\rho}
+ 2 h_{\rho[\nu}R_{\sigma]\mu} \nonumber \\
&+& R h_{\mu[\nu}h_{\sigma]\rho}~, \\
J_{\mu\nu} &=& \frac{1}{3} \left( 2KK_{\mu\sigma}K^{\sigma}_{\nu} +
K_{\sigma\rho}K^{\sigma\rho}K_{\mu\nu} -
2K_{\mu\rho}K^{\rho\sigma}K_{\sigma\nu} \right. \nonumber \\
&-& \left. K^2 K_{\mu\nu} \right)~.
\eea
Two different ways to obtain the junction conditions have been applied so far.
In Ref. \cite{deruelle00} the limit $ \lim_{w \to 0} \int_{-w}^{w}$ of the field equations
(\ref{GB1})-(\ref{GB3}) with the energy-momentum tensor given by (\ref{Tab}) has been taken.
On the other hand, in Ref. \cite{davis} the appropriate Gibbons-Hawking \cite{GH} has been added
to the Gauss-Bonnet action which cancelled normal derivatives of the metric variation.
The form of the boundary term for this theory had been given earlier in Refs. \cite{bunch81,surface}.
In fact, the exact form of the boundary term for the action (\ref{XYZ}) of this paper is not yet known
and according to the claim of Ref. \cite{bunch81} it does not exist.
According to Ref. \cite{Nojiri04} there is even an ambiguity of
the choice of the boundary terms within Gauss-Bonnet brane world. A more general discussion
of the boundary terms is given in Refs. \cite{barvinsky,surface1}.

The discussion of the Gibbons-Hawking boundary term to the
standard first-order Euler density (i.e. Ricci scalar) was also
given, for example, in Ref. \cite{papantopoulos}.

\section{A proposal to formulate consistent junctions conditions in the fourth-order theory
on the brane}
\label{sect3}

\setcounter{equation}{0}

Bearing in mind the roots of the irregularity of the field equations
(\ref{XYZ1})-(\ref{XYZ2}), and using the notation of \cite{israel66}, we
suggest to define the following {\it singular hypersurface of order three} for which the metric derivatives and
the extrinsic curvature on both sides of it are given by
\begin{eqnarray}
\label{hh1}
h^{-}_{\mu\nu} &=& h^{+}_{\mu\nu}~,\\
\label{hh2}
h^{-}_{\mu\nu,w} & = & h^{+}_{\mu\nu,w}~, \hspace{0.5cm} K^{-}_{\mu\nu}
= K^{+}_{\mu\nu}~,\\
\label{hh3}
h^{-}_{\mu\nu,ww} &=& h^{-}_{\mu\nu,ww}~, \hspace{0.5cm} K^{-}_{\mu\nu,w} =
K^{+}_{\mu\nu,w}~,\\
\label{hh4}
h^{-}_{\mu\nu,www} & \neq & h^{+}_{\mu\nu,www}~, \hspace{0.5cm} K^{-}_{\mu\nu,ww}
\neq K^{+}_{\mu\nu,ww}~,
\end{eqnarray}
i.e., the metric and its first derivative are regular, the second derivative of the
metric is continuous, but possesses a kink, the third derivative of the metric
has a step function discontinuity, and as long as the fourth derivative of the
metric on the brane produces the delta function contribution. On the other hand,
according to \cite{israel66} a {\it singular hypersurface of order two} is defined as
\begin{eqnarray}
h^{-}_{\mu\nu} &=& h^{+}_{\mu\nu}~,\\
h^{-}_{\mu\nu,w} & = & h^{+}_{\mu\nu,w}~, \hspace{0.5cm} K^{-}_{\mu\nu}
= K^{+}_{\mu\nu}~,\\
h^{-}_{\mu\nu,ww} &\neq& h^{-}_{\mu\nu,ww}~, \hspace{0.5cm} K^{-}_{\mu\nu,w}
\neq K^{+}_{\mu\nu,w}~,
\end{eqnarray}
and it would describe the boundary surfaces characterized by jumps in the
energy-momentum tensor (e.g. the boundary surface separating a star
from the surrounding vacuum). The physical interpretation of the singular
hypersurfaces of order three is not so obvious, since it should be
characterized by a jump of the first derivative of the
energy-momentum tensor.

\begin{figure}

\begin{center}
        \resizebox{7cm}{!}{
            \includegraphics{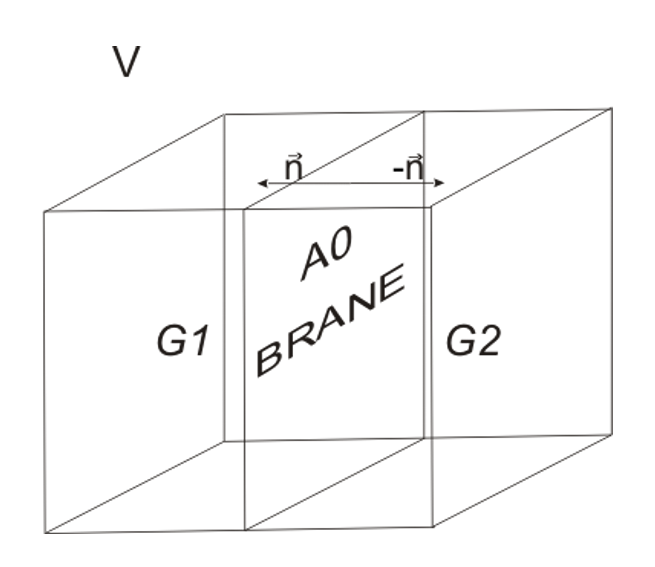}
        }
        \caption[]{A schematic picture illustrating the domains of integration
        used in derivation of the junction conditions. Here $V= G1+G2$,
        $A1= \partial G1 + A0$ and $A2= \partial G2 - A0$.}
        \label{fig1}
    \end{center}
\end{figure}

In order to carry on let us notice that the field equations (\ref{XYZ1})-(\ref{XYZ2})
can be rewritten as
\begin{eqnarray}
\label{Wabd}
\sqrt{-g}C_{ab}{W^{abd}}_{;d} + \sqrt{-g}C_{ab}V^{ab} =
{\chi \over 2} T^{ab}C_{ab}\sqrt{-g}~,
\end{eqnarray}
where we have introduced is an arbitrary tensor field $C_{ab}$, and
\begin{eqnarray}
    W^{abd}&=&f_{X; c }(g^{a b} g^{c d}-g^{(a c} g^{b) d}) + (f_Y
    R^{ab})^{;d} \\ \nonumber
    &+& g^{ab}(f_Y R^{cd})_{;c}  -2 (f_Y R^{d(a})^{;b)}
    - 4(f_Z R^{d(ab)c})_{;c}~,\\
    V^{ab} &=& -\frac{1}{2} f g^{a b} + f_X R^{a b}+2 f_Y R^{c (a}
    {R^{b)}}_{c} \nonumber \\
    &+& 2 f_Z R^{e d c (a} {R^{b)}}_{c d e}~,
\end{eqnarray}
contain third derivatives of the metric. In fact,
${W^{abd}}_{;d}$ is proportional to $\delta(w)$, and the energy-momentum
tensor $T_{ab}$ is given by (\ref{Tab}). Now, we integrate both
sides of (\ref{Wabd}) over the volume $V$ which contains the
following parts (cf. Fig. \ref{fig1}): $G1$, $G2$ - are the
left-hand-side and the right-hand-side bulk volumes which are
separated by the brane, $A1=
\partial G1 + A0$, $A2= \partial G2 - A0$ are the boundaries of
these volumes, and $A0$ is the brane which orientation is given by
the direction of the normal vector $\vec{n}$. We have

\begin{eqnarray}
&& \int_{G1+G2}{\sqrt{-g}C_{ab}{W^{abd}}_{;d} d\Omega}  \\
&+& \int_{G1+G2} {\sqrt{-g}C_{ab}V^{ab}d\Omega}
=\int_{G1+G2}{{\chi \over 2}T^{ab}C_{ab}\sqrt{-g}d\Omega}~, \nonumber
\end{eqnarray}
and so
\begin{eqnarray}
&& \int_{G1+G2}\sqrt{-g}(C_{ab}W^{abd})_{;d} d\Omega
\nonumber \\
&-& \int_{G1+G2}\sqrt{-g}C_{ab;d}W^{abd}
d\Omega  \nonumber \\
&+& \int_{G1+G2}\sqrt{-g}C_{ab}V^{ab}d\Omega  \\ \nonumber
&=& \int_{G1}{\chi \over 2} T^{ab}C_{ab}\sqrt{-g}d\Omega + \int_{G2}{\chi \over 2}
T^{ab}C_{ab}\sqrt{-g}d\Omega \\
\nonumber &+& \int_{A0}{\chi \over 2}
S^{ab}C_{ab}\sqrt{-\gamma}d\sigma~.
\end{eqnarray}
Using the Gauss theorem first
\begin{eqnarray}
    \int_{G1+G2}\sqrt{-g}(C_{ab}W^{abd})_{;d} d\Omega =
    \int_{A1+A2}\sqrt{-\gamma}C_{ab}W^{abd}n_d d\sigma~, \nonumber
\end{eqnarray}
we then integrate in the limit $V\rightarrow A0$ (which in Gaussian normal coordinates
(\ref{bm}) corresponds to the limit $ \lim_{w \to 0} \int_{-w}^{w}$) to obtain
\begin{eqnarray}
    \int_{A0}\sqrt{-\gamma}C_{ab}\{[W]^{abd}n_d - {\chi \over 2} S^{ab}\} = 0~,
\end{eqnarray}
where $[W]^{abd} = W^{abd+} - W^{abd-}$. Since the tensor $C_{ab}$ is arbitrary
one can then conclude that
\begin{eqnarray}
\label{ws}
    [W]^{abd}n_d - {\chi \over 2} S^{ab} &=& 0~,
\end{eqnarray}
and these are exactly the junction conditions for the theory under study.

For example, in $f(X,Y,Z)=f(X)=f(R)$ theory in $D=5$ dimensions with metric
\begin{eqnarray}
\label{bw1}
ds^2&=&-dt^2 \\
&+& a^2(t,w)[dr^2 +r^2(d\Theta^2 +sin^2\Theta d\phi^2)]+dw^2
\nonumber
\end{eqnarray}
they give
\begin{eqnarray}
    [a'''] &=&  {\chi \over 2}{a_0} {p_0}~, \\
    p_0&=& \rho_0~,
\end{eqnarray}
where $(\ldots)' = \partial / \partial w$, $a_0=a(w=0)$ and the
brane energy-momentum tensor $S_{\mu}^{\nu} =
(-\rho_0,p_0,p_0,p_0)$.

\section{Equivalent second-order theory approach}
\label{sect4}
\setcounter{equation}{0}

It is known that the fourth-order gravity theory
\bea
\label{fR}
S &=& \chi^{-1} \int d^D x \sqrt{-g} f(R)
\eea
is equivalent to a second-order theory with the action \cite{f(R)}
\bea
S = \chi^{-1} \int d^D x \sqrt{-g} \left[ f'(Q)\left(Q-R \right)
+ f(Q) \right]
\eea
of which equation of motion is just $Q=R$, provided $f''(Q) \neq
0$. In this approach $f'(Q)$ may be interpreted as an extra scalar
field $\phi=f'(Q)$.

Similar approach can also be used to the fourth-order gravity of
the function of Gauss-Bonnet term
\bea
\label{fRGB}
S &=& \chi^{-1} \int d^D x \sqrt{-g} f(R_{GB})
\eea
by considering the second-order theory action \cite{f(RGB)}
\bea
S = \chi^{-1} \int d^D x \sqrt{-g} \left[ f'(A) \left( R_{GB}-A \right)
+ f(A) \right]~.
\eea
The variation of this action with respect to $A$ gives
the equation of motion $A=R_{GB}$, provided $f''(A) \neq 0$ and $\psi=f'(A)$ is
interpreted as an extra scalar field.

A step ahead can be made by considering more general fourth-order theory
\cite{kijowski,quadratic}
\bea
\label{fRab}
S &=& \chi^{-1} \int d^D x \sqrt{-g} f(g_{ab},R_{ab})
\eea
which can be made equivalent to the second-order theory provided
that one introduces the tensor field
\bea
k^{ab} &=& k^{ab}(g_{ab},R_{ab}) = \frac{\partial f}{\partial
R_{ab}}~,
\eea
which reduces the appropriate field equations to the second-order
and eventually transforms it into the Einstein theory with some new metric tensor.

We follow these considerations by noticing that the theory given by the action
(\ref{XYZ}) is a special case of a
more general theory with the action of the form
\bea \label{r} S_{G} &=& \chi^{-1} \int_{M} d^{D}x \sqrt{-g}
f(g_{ab},R_{abcd})~. \eea
In fact, it is enough to consider the theory (\ref{r}) in order to
get all the previous actions (\ref{fR}), (\ref{fRGB}) and
(\ref{fRab}) as well as the theory with the linear combinations of
curvature terms $f(X,Y,Z) = aX^2 + bY + cZ$, as in Ref. \cite{braneR2}.

In general, the action (\ref{r}) again gives immediately the fourth-order field
equations, but there exist a possibility to formulate it in
an equivalent way by using the action of the following form
\bea
\label{equiv}
 S_{I} &=& \chi^{-1} \int_{M} d^{D}x \sqrt{-g} \{ {\partial
 f(g_{ab},\phi_{abcd}) \over \partial \phi_{ghij}}(R_{ghij}-
 \phi_{ghij}) \nonumber \\
 &+& f(g_{ab},\phi_{cdef}) \}
 \eea
 where $\phi_{abcd}$ is the tensor field which is independent of $g_{ab}$.
 The variation of the action (\ref{equiv}) with respect to $\phi_{abcd}$ gives the
 equation of motion which is simply
 \bea
 \label{1em}
 R_{ghij} = \phi_{ghij},
 \eea
provided that the determinant
\bea
\label{det2}
det \left[{\partial^2
f(g_{ab},\phi_{abcd}) \over \partial \phi_{ghij} \partial \phi_{klmn}} \right]
\neq 0~.
\eea
The condition (\ref{det2}) holds for the fourth-order theory with
the action (\ref{XYZ}), too. This means that (\ref{XYZ}) and
(\ref{equiv}) are the equivalent actions. Introducing the tensor
\bea \label{H} H^{ghij} \equiv {\partial f(g_{ab},\phi_{abcd}) \over
\partial \phi_{ghij}}~, \eea  and then varying the full action
$S=S_{I}+S_{m}$ with respect to both $H^{abcd}$ and $g_{ab}$, we get
the field equations in the following form: \bea \label{em1}
R_{ghij} &=& - {\partial V(g_{ab},H^{cdef}) \over \partial H^{ghij}}~, \\
\label{em2} {1\over 2} g^{ab}f &+& {\partial f \over \partial
g_{ab}} + H^{becd} \phi ^{a}_{~ecd}(g_{ab},H^{klmn})
+ \nonumber \\
&+&\{A^{(ab)cd}\}_{;dc}
 = - {\chi \over 2} T^{ab}~,
\eea
where
\bea
A^{abcd}={1 \over 2} \{H^{acdb} &+& H^{abdc}-H^{cbda} - H^{acbd} + \nonumber \\
&-& H^{abcd}+H^{cbad}\}
\eea
and
\bea
&& V(g_{ab},H^{cdef}) = \\
&-& H^{hgij} \phi_{ghij}(g_{ab},H^{cdef}) +
f(g_{ab},\phi_{klmn}(g_{ab},H^{cdef})) \nonumber
\eea
In fact, the possibility to express the fields $\phi_{abcd}$ as a function of
$g_{ab}$ and $H^{cdef}$ is guaranteed by the condition
(\ref{det2}).

\section{Junction conditions and their correspondence in the equivalent second-order theory}
\label{sect5}

\setcounter{equation}{0}

The tensor
$$ M^{ab} = {1\over 2} g^{ab}f + {\partial f \over \partial g_{ab}} + H^{becd}
\phi ^{a}_{ecd}(g_{ab},H^{klmn})$$ is a function of the fields
$g_{ab}$ and $H^{abcd}$ only, and it does not give any contribution
to the junction conditions. This is because $H^{abcd}$ are
continuous on the brane, their first normal derivatives have a jump,
and their second normal derivatives are proportional to the
$\delta(y)$ function on the brane. Performing the same type of
calculation as in the Section \ref{sect3} the relation (\ref{em2})
gives the junction conditions in the form:
 \bea \label{jc1}
[{A^{(ab)cd}}_{;d}]n_{c} = - {\chi \over 2} S^{ab}~. \eea
Assuming that (\ref{1em}) is fulfilled, and that
\bea \label{f}
f(g_{ab},\phi_{abcd})=f({\phi_{ab}}^{ab},{\phi_{acb}}^{c}{\phi^{acb}}_{c},\phi_{abcd}
\phi^{abcd}),
\eea
we have
\bea \label{equivH}
{[A^{(ab)cd}}_{;d}]n_{c}&=&{[A^{(ab)cd}}_{;c}]n_{d} = [- \{f_{X; c
}(g^{ab} g^{c d}-g^{c(a} g^{b)d}) \nonumber \\  &+& (f_Y  R^{ab})^{;d} + g^{ab}(f_Y R^{cd})_{;c} \\
\nonumber -2(f_YR^{d(a})^{;b)} &-& 4(f_Z R^{d(ab)c})_{;c}\}]n_{d}=
-[W^{abd}]n_{d}. \eea
This means that the junction conditions (\ref{jc1}) in the second-order theory (\ref{equiv}) are completely equivalent to the juction conditions (\ref{ws}) in the fourth-order theory (\ref{XYZ}).

The equation (\ref{em1}), which corresponds to (\ref{1em}) implies
that Riemann tensor is continuous on the brane, since the fields
$\phi_{abcd}$ are regular functions of continuous fields $H^{abcd}$
and $g_{ab}$. This, together with the Gauss-Codazzi equations
(\ref{GC}) imply the following junction conditions: \bea \label{jc2}
[K_{ab}]=0 \\ \nonumber [{{\mathcal{L}}_{\vec{n}} K_{ab}}]=0~, \eea
where ${ \mathcal{L}}_{\vec{n}}$ is a Lie derivative of $H^{a(ef)d}$
along the vector field $\vec{n}$. (Vector field $\vec{n}$ is a
normalized vector field tangent to the Gaussian normal coordinate
$w$). The equations (\ref{jc2}) reconstruct the assumptions
(\ref{jc1}).

Bearing in mind the formulas above, equations (\ref{jc1}) can be
reduced to: \bea \label{jc3}  n_{c} n_{d}
[\mathcal{L}_{\vec{n}} A^{(ab)cd}] = - {\chi \over 2}  S^{ab}~, \eea
where \bea
A^{abcd}&=& f_{X}(g^{ad} g^{cb}-g^{cd} g^{ba})\nonumber \\
\nonumber &+& f_{Y}(2R^{ad}g^{bc} - R^{cd}g^{ba} -R^{ba}g^{cd})\\
&+& 4f_{Z}R^{acbd}~.
\eea

In Gaussian normal coordinates the conditions (\ref{jc2}) and
(\ref{jc3}) read as \bea
[K_{ab,w}]&=&0~,\\
n_{c} n_{d} [A^{(ab)cd}_{~~~~~~~,w}] &=& - {\chi
\over 2} S^{ab}~. \eea In a specific example of $f(X,Y,Z)=Z$ theory
for the metric (\ref{bw1}), the equations (\ref{jc2}) and (\ref{jc3})
give: \bea \label{exjc}
8 {[a'''] \over {a_0}} &=&  {\chi \over 2}{\rho_0}~, \\
{p_0} &=& - {\rho_0}~. \eea

\section{Summary}
\label{sum}
\setcounter{equation}{0}

In this paper we have considered the junction conditions for the
fourth-order brane world given by the action which is a general function of the
curvature invariants $R,R_{ab}R^{ab}$, and $R_{abcd}R^{abcd}$. We imposed the
regularity conditions on the metric tensor which was taken to be of the class $
C^2$ functions of the coordinates. Explicitly, it means that the metric and its first
derivative are regular at the brane, while the second derivative has a kink, the third
derivative of the metric has a step function discontinuity, and the
fourth derivative of the metric gives the delta function
contribution to the field equations. In terms of the seminal notation given
first by Israel \cite{israel66}, these conditions describe the singular
hypersurfaces of order three. The junction conditions which we obtained
are quite generic and they may be applied to some special cosmological
framework of interest.

As an alternative which allows less restrictive regularity
conditions, we considered the reduction of the fourth-order theory to a
second-order theory by applying an extra tensor field. We then
formulated the junction conditions within such a theory and showed
that they were equivalent to the previously obtained fourth-order
theory junction conditions.

In the previously considered cases in the literature mainly the
Eulerian density brane worlds were studied
\cite{deruelle00,charmousis,davis,jim,lidsey,maeda,apostopoulos,meissner01}.
The only non-Eulerian density cases were investigated in Refs.
\cite{branef(R)} and \cite{braneR2}, where the theories with at most
a linear combination the curvature invariants of the form
$f(X,Y,Z) = aX^2 + bY + cZ$ ($a,b,c=$ const.)
were considered. In these references the junction conditions were obtained after transforming
such theories into the equivalent second-order theories.
In fact, we made one step further, by suggesting junction conditions for the
brane world which allows a general
function of curvature invariants $f(X,Y,Z)$. Also, in one of our
approaches to the problem of junction conditions we do not reduce the
fourth-order action into the second-order action, but suggest junction
conditions to work in the fourth-order theory, though in the case of the so-called singular
hypersurfaces of order three only \cite{israel66}.

We hope that our calculations will allow us to study some cosmological
applications of the general fourth-order gravity on the brane (\ref{XYZ}) in the
following papers.

\vspace{0.3cm}

\section{Acknowledgements}


We thank Andrei Barvinsky, Salvatore Capozziello, Tomasz Denkiewicz, Krzysztof Meissner
and David Wands for discussions.
M.P.D. acknowledges the support of the Polish Ministry of Science
and Higher Education grant No 1 P03B 043 29 (years 2005-2007).


\end{document}